\documentclass[twocolumn,pre,aps,showpacs]{revtex4}

\begin{document}

\title{Asymptotic expansion for reversible A $+$ B $\leftrightarrow$ C
       reaction-diffusion process}
\author{Zbigniew Koza}
\email{zkoza@ift.uni.wroc.pl}
\affiliation{Institute of Theoretical Physics,
University of Wroc{\l}aw, plac Maxa Borna 9, PL-50204 Wroc{\l}aw, Poland }

\date{\today}

\begin{abstract}
We study long-time properties of reversible reaction-diffusion systems of
type A $+$ B $\leftrightarrow$ C by means of perturbation expansion in powers
of $1/t$ (inverse of time). For the case of equal diffusion coefficients we
present exact formulas for the asymptotic forms of reactant concentrations
and a complete, recursive expression for an arbitrary term of the expansions.
Taking an appropriate limit we show that by studying reversible reactions
one can obtain ``singular'' solutions typical of irreversible reactions.
\end{abstract}


\maketitle


\section{Introduction}

Behavior of many physical, biological, and chemical systems is determined by
evolution of a reaction front formed between initially separated reactants
\cite{Rice,HavlinBook}. The simplest theoretical model of this phenomenon
consists in assuming that initially two species A and B are uniformly
distributed on opposite sides of an impenetrable barrier.  The barrier  is
removed at time $t = 0$ and the two species start to mix and react, forming
a dynamic reaction front. It is assumed that diffusion is the only transport
mechanism and that the diffusion coefficients of each species are strictly
constant, i.e., independent of spatial location, reactant concentrations,
etc. The problem to be solved is to give a detailed description of
spatiotemporal evolution of this reaction-diffusion system.

Simple as it is,  this  system exhibits many unexpected features,
especially when the reaction is irreversible (A + B $\to$ C). Using a
scaling ansatz, Galfi and R\'acz \cite{Galfi} showed that in this case
the width of the reaction front grows asymptotically as $t^\alpha$
with surprisingly small value of the exponent $\alpha = 1/6$.
Renormalization group technique \cite{LeeFront94}, dimensional
analysis \cite{CornellScaling93,Krapivsky95}, and extensive computer
simulations \cite{CornellSimul} were then employed to demonstrate that
$\alpha = 1/6$ only above the critical dimension $d_c= 2$ and for $d
\le d_c$ one must take into account fluctuations of reactant
concentrations. It was also shown \cite{Ebner,KozaPHA} that the case
where one of the reactants is immobile ($D_{A} = 0$ or $D_{B} = 0$)
belongs to a separate universality class, with $\alpha = 1/2$.
Short-time perturbation \cite{Haim91} and numerical \cite{KozaHaim96}
analysis revealed that the reaction front can move in a nonmonotonic
manner. The quasistatic approximation
\cite{QuasiRedner,CornellScaling93} was used to find a detailed
description of concentrations of species A, B, and C outside the
reaction zone \cite{KozaPHA,KozaJSP,Sinder00} and the case of equal
diffusion coefficient was treated rigorously  \cite{Schenkel93}. These
theoretical results are in full agreement with experiments
\cite{JSPExperim91,HaimExotic,Kopel96,GelExperiment,Leger99} and were
generalized to several more complex reactions
\cite{CornellKoza,Cox01,Yen98,Magnin00,Bazant00,Antal01,SinderPRE02}.

In reality, however, most of chemical reactions are reversible. In spite of
this,
 reversible A
$+$ B $\rightleftharpoons$ C reaction-diffusion processes did not attract so
much attention. This should be probably attributed to the fact that
reversible reaction-diffusion processes for a long time were not supposed to
exhibit any ``anomalous'' properties, especially after Chopard \emph{et.al}
\cite{Chopard93} showed that (a) the front width of a reversible reaction
asymptotically scales with time as if the process was governed only by
diffusion ($w(t) \propto t^{1/2})$ and (b) the fluctuations do not modify
the scaling exponents even in one-dimensional systems. However, the problem
of giving a detailed description of spatiotemporal evolution of reversible
reaction-diffusion systems remained open.

This issue was recently considered by  Sinder and Pelleg in a series of
papers \cite{Sinder99,Sinder00,Sinder00Static}. They restricted their
analysis mainly to systems with very small backward reaction rate $g$ and
found that in this case concentrations of species A, B, and C are
practically the same as those observed in strictly irreversible reactions
($g=0$) everywhere except in a very narrow reaction zone. They confirmed the
result of Ref.\ \cite{Chopard93} that there is a crossover between
intermediate-time ``irreversible'' and long-time ``reversible'' regimes.
They also showed that in contrast to irreversible systems, the effective
asymptotic reaction rate $R$ can have two maxima and there can be even a
region where it is negative. Moreover, they presented strong arguments in
support of a conjecture that reversible reaction-diffusion processes belong
to \emph{two} distinct universality classes. One of them consists of systems
with immobile reaction product C and asymptotically immobile reaction front,
while all other systems belong the other universality class.

The aim of our paper is to work out a new tool for analysis of arbitrary
reversible reaction-diffusion systems---expansion of concentrations as series
in $1/t$ (where $t$ denotes time)---and to use it to find exact solutions of
the problem at least for some particular  combinations of control
parameters. This kind of approach was already successfully applied to
reaction fronts at very short times \cite{Haim91}. To our knowledge,
however, such a technique has not been applied to the long-time regime.

The structure of the paper is as follows. In section \ref{SectionModel} we
give the precise definition of the problem to be solved. In section
\ref{SectionFormalism} we show how to apply the perturbation expansion to
reversible reaction-diffusion systems. In section \ref{SectionEqual} we
employ this technique to study the case where diffusion coefficients of all
species are the same. In particular, we present (a) explicit forms of the
long-time reaction rate $R$ and the concentrations of species A, B, and C;
(b) a reccursive formula for any term of the expansions; (c) detailed
analysis of the crossover to the long-time, reversible regime. In section
\ref{SectionTails} we analyze two particular cases where some aspects of the
system behaviour can be studied analytically. Finally, section
\ref{SectionConclusions} is devoted to discussion of results.


\section{Model}
\label{SectionModel}
 We shall consider the following system of nonlinear
partial differential equations \cite{Galfi}
\begin{eqnarray}
  \label{eq1a}
   \frac{\partial a(x,t)}{\partial t}  &=&
      D_{A} \frac{\partial^2 a(x,t)}{\partial x^2}  -
       R(x,t)\\[1ex]
  \label{eq1b}
   \frac{\partial b(x,t)}{\partial t}  &=&
      D_{B} \frac{\partial^2 b(x,t)}{\partial x^2}  -
       R(x,t)\\[1ex]
  \label{eq1c}
   \frac{\partial c(x,t)}{\partial t}  &=&
      D_{C} \frac{\partial^2 c(x,t)}{\partial x^2}  + R(x,t)
\end{eqnarray}
where
\begin{equation}
   \label{eq1R}
   R(x,t) \equiv k a(x,t) b(x,t) - g c(x,t).
\end{equation}
Here $a(x,t)$, $b(x,t)$, and $c(x,t)$ are concentrations of species A, B and
C, respectively, $D_{A}$, $D_{B}$, and $D_{C}$ are their respective diffusion
coefficients, $R(x,t)$ denotes the effective reaction rate, while $k>0$ and
$g\ge 0$ are the forward and backward reaction rate constants, respectively.
The initial conditions to (\ref{eq1a}) -- (\ref{eq1R}) read
\begin{equation}
\label{iniC}
  a(x,0) = a_0 H(x), \quad
  b(x,0) = b_0 H(-x), \quad c(x,0) = 0,
\end{equation}
where $a_0, b_0$ are initial concentrations of species A and B,
respectively, and $H(x)$ is the Heaviside step function,  which is $0$ for
$x < 0$ and $1$ for $x>0$. Note that in Refs
\cite{Sinder99,Sinder00,Sinder00Static} $R$ was called a ``refined reaction
rate'' and denoted as $R_r$. Our main goal is to give a detailed description
of the long-time solutions of (\ref{eq1a}) -- (\ref{iniC}).

As was pointed out in Ref.\ \cite{Chopard93}, by measuring length, time, and
concentration in units of $\sqrt{D_{A}/ka_0}$, $ 1/k a_0$, and $a_0$,
respectively, our original problem can be reduced to the one with
\begin{equation}
  \label{InitValues}
     D_{A} =1,\quad a_0 = 1, \quad k=1.
\end{equation}
We shall adopt these particular values in our further analysis, which will
leave us with four independent control parameters: $g$, $b_0$, $D_{B}$, and
$D_{C}$.


\section{The asymptotic expansion}
\label{SectionExpansion}

\subsection{Formalism}
\label{SectionFormalism}

The analysis performed in Refs. \cite{Chopard93,Sinder00,Sinder00Static}
shows that in the asymptotic, long-time limit solutions to equations
(\ref{eq1a}) - (\ref{eq1R}) take on a scaling form
\begin{eqnarray}
  \label{s1a}
   a(x,t) &\approx& S_{A}(x/\sqrt{t}), \\[1ex]
  \label{s1b}
   b(x,t) &\approx& S_{B}(x/\sqrt{t}), \\[1ex]
  \label{s1c}
   c(x,t) &\approx& S_{C}(x/\sqrt{t}), \\[1ex]
  \label{s1R}
   R(x,t) &\approx& t^{-1}S_{R}(x/\sqrt{t}).
\end{eqnarray}
We therefore assume that in the long-time limit all functions  involved in
(\ref{eq1a}) -- (\ref{eq1R}) can be expanded as series in $\tau \equiv 1/t$,
with coefficients being some functions of $\xi \equiv x/\sqrt{t}$. We thus
write $a(x,t)$, $b(x,t)$, and $c(x,t)$ as
\begin{eqnarray}
  \label{expA}
   a(x,t) &=& \sum_{n=0}^{\infty} \tau^n {\cal A}_n(\xi), \\[1ex]
  \label{expB}
   b(x,t) &=& \sum_{n=0}^{\infty} \tau^n {\cal B}_n(\xi), \\[1ex]
  \label{expC}
   c(x,t) &=& \sum_{n=0}^{\infty} \tau^n {\cal C}_n(\xi).
\end{eqnarray}
Using  (\ref{eq1R}) we conclude that $R(x,t)$ can be expressed as
\begin{equation}
 R(x,t) = \sum_{n=0}^{\infty} \tau^n {\cal R}_n(\xi),
\end{equation}
where
\begin{equation}
  \label{expR}
    {\cal R}_n(\xi) \equiv   - g{\cal C}_n(\xi)
                        + \sum_{j=0}^{n}
                             {\cal A}_j(\xi){\cal B}_{n-j}(\xi).
\end{equation}

Substituting (\ref{expA}) -- (\ref{expR}) into (\ref{eq1a}) --
(\ref{eq1R}) and collecting coefficients at $\tau^0$
we find that
\begin{equation}
\label{R0=0}
   {\cal R}_0(\xi) \equiv {\cal A}_0(\xi){\cal B}_0(\xi) -
                             g{\cal C}_0(\xi) = 0.
\end{equation}
This equation expresses a fundamental property of the system: in the
long-time limit it tends to a local chemical equilibrium (the forward
and backward reaction rates become asymptotically equal)
\cite{Chopard93}. Collecting now coefficients at $\tau^{n+1}$,
$n=0,1,\ldots$, we arrive at
\begin{eqnarray}
     {\cal A}''_n(\xi) +  \frac{1}{2} \xi {\cal A}'_n(\xi) +
        n {\cal A}_n -
    {\cal R}_{n+1}(\xi) &=& 0 \\[1ex]
    D_{B} {\cal B}''_n(\xi) + \frac{1}{2} \xi {\cal B}'_n(\xi) +
       n {\cal B}_n - {\cal R}_{n+1}(\xi) &=& 0 \\[1ex]
 \label{Eq-dc}
    D_{C} {\cal C}''_n(\xi) + \frac{1}{2} \xi {\cal C}'_n(\xi) +
       n {\cal C}_n
       + {\cal R}_{n+1}(\xi) &=& 0
\end{eqnarray}
where we used a short-hand notation $f'(\xi) \equiv {d} f/{d}\xi$, $f''(\xi)
\equiv {d}^2 f/{d}\xi^2$. In the lowest order ($n=0$) we thus find
\begin{eqnarray}
  \label{pdfA}
   {\cal A}''_0(\xi) + \frac{1}{2} \xi {\cal A}'_0(\xi)
         &=& {\cal R}_1(\xi) \\[1ex]
  \label{pdfB}
   D_{B} {\cal B}''_0(\xi) + \frac{1}{2} \xi {\cal B}'_0(\xi)
         &=& {\cal R}_1(\xi) \\[1ex]
  \label{pdfC}
   D_{C} {\cal C}''_0(\xi) + \frac{1}{2} \xi {\cal C}'_0(\xi)
         &=& -{\cal R}_1(\xi).
\end{eqnarray}
Equations (\ref{R0=0}), (\ref{pdfA}) -- (\ref{pdfC}) constitute a system of
four equations for four unknown functions: ${\cal A}_0(\xi)$, ${\cal
B}_0(\xi)$, ${\cal C}_0(\xi)$, and ${\cal R}_1(\xi)$. These important
functions control the asymptotic (long-time) properties of the system and
can be readily identified with the scaling functions employed in relations
(\ref{s1a}) -- (\ref{s1R}). However, owing to a nonlinear form of the
equation (\ref{R0=0}), the explicit form of these functions can be found
only in a few particular cases discussed below.



\subsection{The case of equal diffusion constants}
\label{SectionEqual}
\subsubsection{Asymptotic solution}

Following (\ref{InitValues}) we will now assume $D_{A} = D_{B} = D_{C} = 1$,
$k=1$ and $a_0=1$. The only free parameters of the problem are thus $b_0$
and $g$. Upon adding (\ref{eq1c}) to (\ref{eq1a}) and (\ref{eq1b})  we
arrive at two diffusion equations with well known solutions
\begin{eqnarray}
  a(x,t) + c(x,t) &=& \frac{1}{2}{\,\rm erfc}(x/\sqrt{4t}),\\[1ex]
  b(x,t) + c(x,t) &=& \frac{1}{2}b_0{\,\rm erfc}(-x/\sqrt{4t}).
\end{eqnarray}
This immediately leads to
\begin{eqnarray}
  \label{solA0}
  {\cal A}_0(\xi) &=&
     \frac{\Phi(\xi) - g + \sqrt{\Delta(\xi)}}{2}
  \\[1ex]
  \label{solB0}
  {\cal B}_0(\xi) &=&
     \frac{-\Phi(\xi) - g + \sqrt{\Delta(\xi)}}{2}
  \\[1ex]
  \label{solC0}
  {\cal C}_0(\xi) &=&
     \frac{{\,\rm erfc}(\xi/2) + b_0{\,\rm erfc}(-\xi/2) + 2g
        -2\sqrt{\Delta(\xi)}}{4}
  \\[1ex]
  \label{solR1}
  {\cal R}_1(\xi) &=&
     \frac{g b_0 (1 + g + b_0) \exp(-\xi^2/2)}{2\pi[\Delta(\xi)]^{3/2}} \\[1ex]
  \label{solABCn}
  {\cal A}_n(\xi) &=&
      {\cal B}_n(\xi) = -  {\cal C}_n(\xi), \quad n \ge 1
\end{eqnarray}
where we denoted
\begin{eqnarray}
 \label{DefPhi}
   \Phi(\xi) &\equiv&  \frac{1}{2}[a_0 {\,\rm erfc}(\xi/2) - b_0  {\,\rm erfc}(-\xi/2)]
 \\[1ex]
 \label{DefDelta}
\Delta(\xi) &\equiv&
                \left[
                        \Phi(\xi) -  g
                \right]^2
                    + 2 g {\,\rm erfc}(\xi/2).
\end{eqnarray}

Note that ${\cal R}_1(\xi)$ can diverge to infinity. This can happen if and
only if $\Delta(\xi) \to 0$ which, in turn, occurs only if $\Phi(\xi) \to 0$
and $g\to 0$.
 Since $\Phi(\xi)$
decreases monotonically from $a_0$ to $-b_0$, equation $\Phi(\xi) = 0$ has a
unique solution, which will be denoted $\xi_f$.

In the limit of an irreversible reaction we find
\begin{eqnarray}
 \label{g_to_0_A}
     \lim_{g\to 0} {\cal A}_0(\xi) &=&
            \left\{
                \begin{array}{ll}
                    \Phi(\xi), & \xi < \xi_{f} \\
                    0, & \xi \ge \xi_{f}
                \end{array}
            \right.
  \\[1ex]
 \label{g_to_0_B}
     \lim_{g\to 0} {\cal B}_0(\xi) &=&
            \left\{
                \begin{array}{ll}
                    0, & \xi < \xi_{f} \\
                    -\Phi(\xi), & \xi \ge \xi_{f}
                \end{array}
            \right.
  \\[1ex]
 \label{g_to_0_C}
     \lim_{g\to 0} {\cal C}_0(\xi) &=&
            \left\{
                \begin{array}{ll}
                    \frac{1}{2} b_0{\,\rm erfc}(-\xi/2), & \xi < \xi_{f} \\
                    \frac{1}{2} {\,\rm erfc}(\xi/2), & \xi \ge \xi_{f}
                \end{array}
            \right.
  \\[1ex]
 \label{g_to_0_R}
     \lim_{g\to 0} {\cal R}_1(\xi) &=&
      \frac{b_0 \exp(-\xi_{ f}^2/4)}{\sqrt{\pi}{\,\rm erfc}(\xi_{ f}/2)}
      \delta(\xi - \xi_{ f})
\end{eqnarray}
where $\delta$ is the Dirac's delta distribution.  These relations are in
full agreement with general formulas derived in Ref.\ \cite{KozaJSP} for the
case of a strictly irreversible reaction, $g=0$, and in Ref.\
\cite{Sinder99} for $a_0 = b_0$ and $g\to 0$. Both species, A and B, become
effectively segregated at $\xi_{f}$. We can thus identify $\xi_f$ as the
position of the reaction front. Beyond this point the reaction rate tends to
$0$ and the concentrations $a(x,t)$, $b(x,t)$, and $c(x,t)$ satisfy
diffusion equations (\ref{eq1a}) -- (\ref{eq1c}) with $R(x,t) = 0$. However,
the existence of a singularity in the above formulas does not mean that the
reaction should be asymptotically restricted to
 a single point!
Recalling that $\xi \equiv x/\sqrt{t}$ we conclude that this singularity
indicates that the reaction must be restricted to a region much narrower than
$\sqrt{t}$; it also suggests that for $g=0$ and $x/\sqrt{t} \approx \xi_{
f}$ we should try and take into account higher-order terms of expansions
(\ref{expA}) -- (\ref{expC}).

In the oposite limit of an infinitely large backward reaction rate $g$ we find
\begin{eqnarray}
 \label{g_to_infA}
     \lim_{g\to \infty} {\cal A}_0(\xi) &=& \frac{1}{2} {\,\rm erfc}(\xi/2)\\[1ex]
 \label{g_to_infB}
     \lim_{g\to \infty} {\cal B}_0(\xi) &=& \frac{1}{2} b_0{\,\rm erfc}(-\xi/2)\\[1ex]
 \label{g_to_infCR}
     \lim_{g\to \infty} {\cal C}_0(\xi) &=& \lim_{g\to \infty} {\cal R}_1(\xi) = 0.
\end{eqnarray}
These equations express the fact that for large backward reaction rates $g$
if a particle C is being created as a result of a forward, A $+$ B $\to$ C
reaction, it is being immediately converted back into a pair A-B;
consequently, the concentration of particles C  tends to 0 and the
concentrations of particles A and B  evolve as if there was no reaction at
all.

\subsubsection{Recursive formula}

To find the remaining terms of expansions (\ref{expA})--(\ref{expC}) we
employ relations (\ref{expR}), (\ref{Eq-dc}), (\ref{pdfC}), and
(\ref{solABCn}), arriving at a recursive formula
\begin{equation}
 \label{Recursive}
 {\cal C}_{n+1}(\xi) = \frac{
      n {\cal C}_n(\xi) + \frac{1}{2} \xi {\cal C}'_n(\xi) +
      {\cal C}''_n(\xi) + S_n(\xi)
                            }{\sqrt{\Delta(\xi)}},
\end{equation}
where $S_n(\xi) \equiv \sum_{j=1}^{n}{\cal C}_j(\xi){\cal C}_{n+1-j}(\xi)$
and $n\ge 0$. An important feature of this relation is that it allows to
express the $(n+1)$-th term of the expansion directly as a function of
already determined, lower-order terms. Consequently, together with the
explicit form of the zeroth order terms given in (\ref{solA0}) --
(\ref{solR1}), equations  (\ref{solABCn}) and (\ref{Recursive}) enable one
to calculate (at least in principle) an arbitrary term of the expansions
(\ref{expA}) -- (\ref{expC}) analytically. In practice, however, the
complexity of the appropriate formulas grows rapidly and even using
computer-algebra systems it is very difficult to determine ${\cal C}_n(\xi)$
for more than a few smallest values of $n$. Notice also that for $\xi=\xi_f$
and $g\to 0$ the denominator of (\ref{Recursive}) goes to 0.

\subsubsection{Crossover from ``irreversible'' to ``reversible'' reaction
fronts }

Let $t^*$ denote the time when the system enters the asymptotic, long-time
time regime. For times $t \gg t^*$ each sum in expansions (\ref{expA}) --
(\ref{expC}) should be dominated by its lowest-order nonvanishing term,
while for $t \alt t^*$ the system should behave as if the reaction was
strictly irreversible ($g = 0$) \cite{Chopard93,Sinder00}. This cross-over
time can be estimated from a relation ${\cal A}_0(\xi_{ f}) \approx {\cal
A}_1(\xi_{ f})/t^*$ . Using (\ref{pdfC}), (\ref{solABCn}), and
(\ref{Recursive}) we find ${\cal A}_1(\xi) = -{\cal C}_1(\xi) = {\cal
R}_1(\xi)/\sqrt{\Delta(\xi)}$. In the limit $g \to 0$ there is thus ${\cal
A}_0(\xi_{ f}) \propto \sqrt{g}$ and ${\cal A}_1(\xi_{ f}) \propto 1/g$.
Therefore in this limit we have
\begin{equation}
\label{t^*}
  t^* \propto g^{-3/2}.
\end{equation}

Note that Chopard \emph{et.al.}\ \cite{Chopard93} proposed a different
relation, $t^* \propto g^{-1}$. However, their conjecture was based on
numerical analysis of a reaction-diffusion system with a very specific
choice of system parameters: $a_0 = b_0$, $D_{A} = D_{B}$, and $D_{C} = 0$.
In other words, they investigated only symmetric systems with immobile
particles C. Moreover, they assumed that the width of the reaction front can
be identified width the width of the profile of particles C, $w_C(t) \equiv
\int x^2c(x,t)/\!\int c(x,t) dx$. Actually this is acceptably only when both
particles C and the reaction front center are immobile ($D_{C} = 0$, $x_f(t)
\sim 0$), a condition implicitly satisfied in their simulations. For $D_{C}
\neq 0$ or $x_f(t) \neq 0$ we expect that asymptotically $w_C(t) \propto
t^{1/2}$, while $w(t) \propto t^{1/6}$ (at least for $g=0$ \cite{Galfi}), so
$w_c(t)$ cannot be identified with $w(t)$. From this point of view the case
studied numerically in Ref.\ \cite{Chopard93} belongs to a separate
universality class. To the same conclusion, though on different grounds,
came Sinder and Pelleg \cite{Sinder00Static}, who investigated systems with
immobile reaction product ($D_{C} =0$). They found that asymptotically
$w(t,g) \propto g^{1/2}t^{1/2}$ for systems with the moving reaction front
($x_f \neq 0$) and $w(t,g) \propto g^{1/3}t^{1/2}$ if $x_f \sim 0$.
Therefore, on taking into account that for small $g$ and intermediate times
$t$ one expects $w(t,g) \propto g^0 t^{1/6}$, we immediatelly arrive at the
conclusion that for asymptotically immobile reaction fronts studied in Ref.\
\cite{Chopard93} there should be $t^* \propto g^{-1}$, while for systems
with $D_{C} = 0$, $g \ll 1$, and a mobile reaction front ($x_f \neq 0$) the
crossover time is given by (\ref{t^*}).

We verified the validity of relation (\ref{t^*}) for systems studied in this
section by computer-assisted analysis of ${\cal A}_n(\xi_{ f})$ for the
fully symmetric case  $a_0 = b_0 =1$. It indicates that
\begin{equation}
   {\cal A}_n(\xi_{ f})  \propto (-1)^n {g^{(1-3n)/2}}
\end{equation}
for all $n \le 5$. Most probably this relation continues to be true also for
higher-order terms. This would mean that for $t \gg t^*$, as expected,
expansions (\ref{expA}) -- (\ref{expC}) will be dominated by their first
terms, while for $t \ll t^*$ they are divergent at $\xi_f$.

Using (\ref{t^*}) we can estimate the width $w$ and height $h$ of
 the reaction front
at the cross-over time. The former is defined as a square root of
$\int_{-\infty}^\infty (x-x_{ f})^2 R(x,t) { d} t /
 \int_{-\infty}^\infty R(x,t)
{ d} t$, while $h = R(x_{ f}, t)$ (here $x_{ f} \approx \sqrt{t}\xi_{ f}$ is
the exact location of the reaction front, see Ref.\ \cite{Galfi,KozaJSP}).
Using (\ref{solR1}) and (\ref{g_to_0_R}) we find
\begin{eqnarray}
 \label{w}
    w(t^*) &\propto& \sqrt{g t^*} \propto (t^{*})^{1/6}, \\[1ex]
 \label{h}
     h(t^*) &\propto& (t^*)^{-1}/\sqrt{g} \propto (t^{*})^{-2/3}.
\end{eqnarray}
These are the scaling relations derived by Galfi and R\`acz for the case of
a strictly irreversible reaction, $g=0$ and $t \to \infty$ \cite{Galfi}.


\subsection{Other cases with rigorous solutions}
\label{SectionTails}

\subsubsection{The limit $|\xi| \to \infty$, or the tails of the
distributions}

Let us now consider the general case of arbitrary values of parameters
$b_0$, $D_{B}$, $D_{C}$ and $g$ in  the limit of $t\to \infty$ and
$|x|\to\infty$ such that  $|x|/\sqrt{t} = |\xi| \to \infty$. For
sufficiently large $\xi$ we may expect that the concentration of particles B
will be very close to its original value $b_0$. Substituting ${\cal
B}_0(\xi) \approx b_0$ in (\ref{pdfA}) -- (\ref{pdfC}) we find that $ {\cal
R}_1(\xi) \approx 0$ and
\begin{equation}
  \label{BigXi+}
    {\cal C}_0(\xi) \approx   b_0g^{-1}{\cal A}_0(\xi)
    \approx \eta^+ {\,\rm erfc} \! \left(
                                 \xi/\sqrt{4 D_{\rm eff}^+}
                             \right)
\end{equation}
where
\begin{equation}
  \label{defDeff+}
  D_{\rm eff}^+ \equiv \frac{g D_{A} + b_0 D_{C}}{g + b_0}
\end{equation}
and $\eta^+$ is an integration constant independent of $\xi$. Note that, as
might be expected on physical grounds, $D^+_{\rm eff}$ lies between $D_{A}$
and $D_{C}$.

Similarly, in the opposite limit $\xi\to-\infty$, we assume
${\cal A}_0(\xi) \approx a_0 = 1$ and find
\begin{equation}
  \label{BigXi-}
    {\cal C}_0(\xi) \approx g^{-1}{\cal B}_0(\xi)
       \approx \eta^-
                      {\,\rm erfc}  \!\left(
                                    -\xi/\sqrt{4 D_{\rm eff}^-}
                                \right)
\end{equation}
where
\begin{equation}
  \label{defDeff-}
  D_{\rm eff}^- \equiv \frac{g D_{B} + D_{C}}{g + 1}
\end{equation}
and $\eta^-$ does not depend on $\xi$. We thus see that sufficiently
 far away from the reaction region all the species diffuse with
 an effective diffusion coefficient $D^+_{\rm eff}$ (for $\xi \to \infty$)
or $D^-_{\rm eff}$ (for $\xi \to -\infty$). Note that $D^+_{\rm eff}$,
$D^-_{\rm eff} \to D_{C}$ as $g \to 0$, in agreement with the findings of
Ref. \cite{Sinder00}.

\subsubsection{Reaction front at $\xi = 0$ for one or two vanishing diffusion constants}

Equations (\ref{pdfB}) -- (\ref{pdfC}) immediately imply that if $D_{B}$ or
$D_{A}$ vanishes then
\begin{equation}
   {\cal R}_1(0) = 0 .
\end{equation}
 Thus either ${\cal R}_1(\xi)$ has a local minimum at
$\xi = 0$ or it attains \emph{negative} values in the vicinity of $\xi = 0$.
Both possibilities have actually been observed in numerical simulations
carried out in Ref.\ \cite{Sinder00Static}.


\section{Conclusions}
\label{SectionConclusions}
We have applied perturbation analysis to reversible reaction-diffusion
systems with arbitrary values of control parameters. Using this technique we
obtained a system of four equations completely governing the long-time
behaviour of the system. We then concentrated on several cases where
analytical results can be derived.

In particular, we found the complete solution of the problem for the case
where the diffusion coefficients of all species are the same. Its most
interesting feature is the limit of vanishingly small backward reaction rate
constant $g$. In this limit, as expected, the solutions become singular at a
point which can be identified with the  reaction-front center. Our method
enables one to analyze explicitly how these singularities, especially the
Dirac's delta function in the expression for the local reaction rate ${\cal
R}_1$, appear in mathematical formulas. This holds out hope that it will be
possible to construct a unified theory of reversible and irreversible
reaction-diffusion systems. It should be also noticed that our asymptotic
solution is also valid for $g=0$ even though in this case all higher-order
terms of the expansion diverge.

We also found a general recursive formula for any term of the expansion as a
function of all lower-order terms. Unfortunately, complexity of expressions
thus obtained grows very rapidly. Nevertheless we believe that this formula
opens a new way of analyzing the evolution of reaction-diffusion systems at
arbitrary times.

Using the information about the first correction to the asymptotic solution
we showed that the anomalous properties of irreversible reaction-diffusion
systems can be studied by taking a suitable limit in the formulas obtained
for reversible systems. A potential advantage of this approach is that
mathematical description of reversible reaction-diffusion systems is more
regular, and hence more amenable to rigorous analysis.

We also studied the crossover time $t^*$ between intermediate-time
(``irreversible reaction'') and long-time (``reversible reaction'') regimes.
We proved that in the case of equal diffusion coefficients $t^*$ scales
with  $g$ as $g^{-3/2}$. This conclusion agrees with results obtained by
Sinder and Pelleg \cite{Sinder00Static} and disagrees with those obtained by
Cornell and Droz \cite{CornellScaling93}. This discrepancy can be easily
understood if one notices that Cornell and Droz studied only systems where
the width $w_c$ of the concentration of species C grows in time as
$t^{1/6}$, while our study was performed for a system where $w_c \propto
t^{1/2}$.

The main message coming from our work may be summarized as follows: (a)
reversible reaction-diffusion systems are more amenable to rigorous
treatment than irreversible ones; (b) it is possible to investigate the more
difficult irreversible reaction-diffusion systems by taking an appropriate
limit in the reversible ones; (c) investigation of reversible
reaction-diffusion systems is interesting not only \emph{per se}, but
constitutes an alternative technique of analyzing irreversible systems.

\bibliographystyle{apsrev}

\end{document}